\documentclass[10pt,a4paper]{memoir}

\usepackage[british]{babel}
\usepackage[sort]{cite}

\usepackage{graphicx}
\usepackage{amsmath}
\usepackage{authblk}
\usepackage{url}
\usepackage[colorlinks=true,allcolors=blue,bookmarks=false]{hyperref}

\usepackage{tjp-layout-article}

\newcommand{\tjpstitle}{Measurement of optical to electrical and electrical to optical delays with ps-level uncertainty}

\newcommand{\tjpsfooter}{\rule{\textwidth}{0.5pt}\par\small\tjpstitle\\\raisebox{-4pt}{\includegraphics[height=10pt]{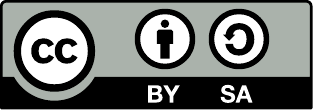}}\hspace{5pt}{\tiny This work is licensed under the \href{http://creativecommons.org/licenses/by-sa/4.0/}{Creative Commons Attribution-ShareAlike 4.0 International License.}}}

\makeevenhead{tjpheader}{}{}{}
\makeoddhead{tjpheader}{}{}{}
\makeevenfoot{tjpheader}{\thepage}{\tjpsfooter}{}
\makeoddfoot{tjpheader}{}{\tjpsfooter}{\thepage}

\graphicspath{{./figures/}}

\begin{document}

\title{\tjpstitle}

\author[1,*]{H.~Z.~Peek}
\author[2,1,3]{T.~J.~Pinkert}
\author[1]{P.~P.~M.~Jansweijer}
\author[2]{J.~C.~J.~Koelemeij}

\affil[1]{Nikhef, Science Park 105, 1098 XG Amsterdam, The Netherlands}
\affil[2]{Department of Physics and Astronomy, LaserLaB, VU University, De Boelelaan 1081, 1081 HV Amsterdam, The Netherlands}
\affil[3]{now at: Physikalisch-Technische Bundesanstalt, Bundesallee 100, 38116 Braunschweig, Bundesrepublik Deutschland}
\affil[*]{Corresponding author: henkp@nikhef.nl}

\maketitle

{\vfill}

\section*{Copyright}
Copyright (C) 2018\\

\noindent{}Vrije Universiteit (Tjeerd J. Pinkert, Jeroen C. J. Koelemeij),\\
Nikhef (Peter P. M. Jansweijer, Tjeerd J. Pinkert),\\
Henk Z. Peek,\\
Tjeerd J. Pinkert,\\
Jeroen C. J. Koelemeij\\

\noindent{}\includegraphics{by-sa}

\noindent{}This work is licensed under the Creative Commons Attribution-ShareAlike 4.0 International License. To view a copy of this license, visit \href{http://creativecommons.org/licenses/by-sa/4.0/}{http://creativecommons.org/licenses/by-sa/4.0/} or send a letter to Creative Commons, PO Box 1866, Mountain View, CA 94042, USA.

\newpage
\begin{center}
{\LARGE \tjpstitle}
\end{center}

\begin{abstract}
We present a new measurement principle to determine the absolute time delay of a waveform from an optical reference plane to an electrical reference plane and vice versa.
We demonstrate a method based on this principle with 2 ps uncertainty.
This method can be used to perform accurate time delay determinations of optical transceivers used in fibre-optic time-dissemination equipment.
As a result the time scales in optical and electrical domain can be related to each other with the same uncertainty.
We expect this method to break new grounds in high-accuracy time transfer and absolute calibration of time-transfer equipment.
\end{abstract}

\section{Introduction}
Relative time delays corresponding to optically generated frequencies can be measured electronically with zeptosecond precision~\cite{lit:pap:NPho-xie-2017}.
It has been shown that electrical to electrical (EE) and optical to optical (OO) time delays can be determined with picosecond level precision~\cite{lit:pap:RSI-prochazka-2012,lit:pap:OE-sotiropoulos-2013}.
However, the absolute time delay between an optical reference plane and an electrical reference plane is much harder to measure.
At best one can estimate the optical to electrical (OE) delay e.g. from the detector geometry and electrical properties, with uncertainties as low as $\pm 7$ ps reported~\cite{lit:pap:M-prochazka-2011}.
Modern opto-electronic time distribution systems, such as White Rabbit (WR)~\cite{lit:proc:proc-serrano-2009}, have become widely used and achieve time interval measurements with picosecond-level precision~\cite{lit:proc:proc-serrano-2009,lit:pap:M-kodet-2016}.
A method to directly measure electrical to optical (EO) delay with ps precision has so far been lacking, but is urgently needed to calibrate such systems.

In this publication we present a new method to characterize OE and EO time delay with respect to defined reference planes.
Although the research described in this article was initiated by the demand for WR absolute calibration~\cite{lit:misc:wr_abs_calibration}, it enables the absolute calibration of any time transfer system that uses optoelectronic conversions.
The presented principle is expected to be generally useful for time and frequency metrology.

\section{Theory}
Optical communication typically involves modulation of the amplitude, phase, or combinations thereof.
Often, amplitude modulation (AM) is used.
For delay calibration method reported here we use a LiNbO3 Mach-Zehnder amplitude modulator (MZM).
In this device an incident optical field is split and sent into two parallel arms of an optical waveguide interferometer.
In one of the arms the optical field interacts with a radio frequency (RF) electric field, co-propagating with the optical field at approximately the same velocity, resulting into nearly instantaneous modulation of the optical field's phase through the electro-optic (Pockels) effect.
At the end of the interferometer the two optical waves are recombined, leading to amplitude modulation of the emanating optical field.
Note that the MZM is essentially a four-port device with two optical and two electrical ports, corresponding to the ends of a bidirectional optical and a bidirectional electrical waveguide, respectively.
A photo diode receiver detects the amplitude modulation of the optical wave and converts it back to the electrical domain.
The nearly instantaneous interaction in the MZM from the electrical to the optical wave is used to accurately determine the time delay between the optical reference plane and the electrical reference plane of the opto-electrical receiver.

\begin{figure}[htbp]
    \centering
    \fbox{\includegraphics[scale=0.9]{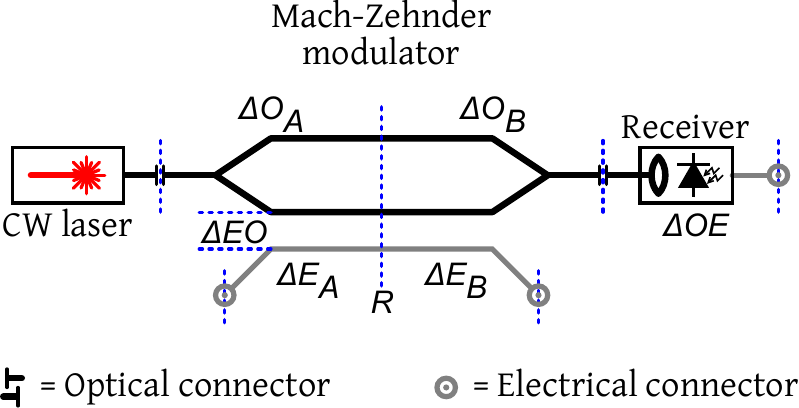}}
    \caption{Proposed experimental set-up for the determination of the optical to electrical delay $\Delta OE$. The reference planes for the time delay determination are indicated by the blue dashed lines.}\label{fig:measurement_setup}
\end{figure}

Figure~\ref{fig:measurement_setup} shows the schematic of the proposed set-up to determine the OE delay, $\Delta OE$, of a receiver.
The measurable delays $D_{1..4}$ between the various reference planes in this device are given by
\begin{subequations}
\begin{align}
    D_{1} &= \Delta E_{A} + \Delta E_{B},\label{eq:delay_ee}\\
    D_{2} &= \Delta O_{A} + \Delta O_{B},\label{eq:delay_oo}\\
    D_{3} &= \Delta E_{A} + \Delta O_{B} + \Delta EO + \Delta OE,\label{eq:delay_eo_a}\\
    D_{4} &= \Delta E_{B} + \Delta O_{A} + \Delta EO + \Delta OE,\label{eq:delay_eo_b}
\end{align}
\end{subequations}
where $\Delta E_{A}, \Delta E_{B}, \Delta O_{A}, \Delta O_{B}$ are delays from the four MZM ports to a virtual reference plane $R$, and $\Delta EO, \Delta OE$ are internal EO and OE delays of respectively the MZM and opto-electrical receiver.
Combining~\eqref{eq:delay_ee} to~\eqref{eq:delay_eo_b} yields
\begin{equation}
    \Delta EO + \Delta OE = \frac{D_{3} + D_{4} - D_{1} - D_{2}}{2}\label{eq:main_oe}
\end{equation}
Assuming that de delays $D_{1..4}$ can be measured, and under the assumption that $\Delta EO$ is very small and can be measured or estimated from the physical dimensions and composition of the MZM traveling wave device, $\Delta OE$ can be determined from this equation.
Its uncertainty is determined by measurement uncertainties and the uncertainty in $\Delta EO$.
Once calibrated, the known delay $\Delta OE$ of the receiver can be used to determine an unknown EO delay $\Delta EO_{\text{t}}$ of a transmitter, by direct EE delay measurement of $\Delta EO_{\text{t}} + \Delta OE$.

\section{Method}
EE and OO delays can be determined with high precision using interpolated cross-correlation ~\cite{lit:pap:ITSP-jacovitti-1993,lit:pap:DSP-zhang-2006,lit:pap:OE-sotiropoulos-2013} (see Supplementary material).
First the delay between the input and output signals of a reference measurement chain is determined.
Then a Device Under Test (DUT) is inserted into the reference measurement chain.
The DUT delay is determined by the difference between the measured delay and the delay of the reference measurement chain.

Fast random signals are advantageous for delay estimation with high time resolution through cross correlation.
In our experiment a 1.25 Gbit/s PRBS-7 signal\footnote{Psuedo Random Binary Sequence (PRBS)} is used.
The choice for the PRBS-7 signal was made because it has similar power spectral density and group delay as that of the IEEE 802.3-BX gigabit Ethernet signal used in the time transfer systems to be calibrated.
Given the fact that a coarse estimation can be made from the expected delay with sufficient accuracy, the correct correlation peaks can always be found within a single PRBS-7 sequence period (T = 101.6 ns).

For the measurement of the delays defined by~\eqref{eq:delay_eo_a} and~\eqref{eq:delay_eo_b}, both the optical and electrical wave guides need to be bi-directional.
This is in principle the case for every MZM.
However, most packaged MZMs have an internally terminated electrical transmission line that only allows a single measurement (either  $D_{3}$ or $D_{4}$).
For this experiment a MZM was modified.
The internal termination was removed from the electrical wave guide, and replaced by an SMA connector.
The resulting device allows us to make the proposed measurements.
The modified MZM has two female electrical connectors (SMA) and two optical connectors (FC/APC).

Suitable connectors need to be selected because proper impedance matching is a requirement.
SMA and 3.5 mm connectors are found to have sufficient connection repeatability for time transfer purposes with picosecond precision~\cite{lit:misc:pinkert-2016}.
The electrical reference planes for the measurements are taken at the connectors as specified in MIL-STD-348B.
The optical reference planes are taken at the physical contact plane of the FC/APC connectors.

\begin{figure}[htbp]
       \centering
       \fbox{\includegraphics[scale=1.8]{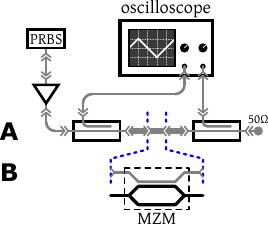}}
       \caption{\textbf{A}) electrical delay chain reference measurement (= electrical reference delay \textit{plus} the delay of an \hbox{F--F} coupling device). \textbf{B}) arrangement used to determine the MZM EE delay $D_{1}$ using reference measurement \textbf{A}.}\label{fig:mzm_electrical_delay_measurement}
\end{figure}

\begin{figure}[htbp]
       \centering
       \fbox{\includegraphics[scale=1.8]{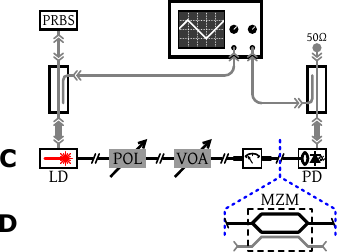}}
       \caption{\textbf{C}) optical delay chain reference measurement (= optical reference delay). \textbf{D}) determine OO delay $D_{2}$ using reference measurement of \textbf{C}.}\label{fig:mzm_optical_delay_measurement}
\end{figure}

\begin{figure}[htbp]
       \centering
       \fbox{\includegraphics[scale=1.8]{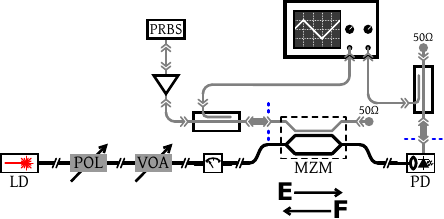}}
       \caption{\textbf{E}) Determination of the EO delay $D_{3}$ (i.e. MZM delays plus $\Delta OE$ photo diode). \textbf{F}) Delay $D_{4}$ is determined by exchange of the MZM ports such that the electrical signal and light travel in the reverse direction through the MZM.}\label{fig:mzm_electrical_optical}
\end{figure}

Device delays are measured as outlined in Figs.~\ref{fig:mzm_electrical_delay_measurement} through~\ref{fig:mzm_electrical_optical}\footnote{Symbols used throughout the figures:
       
\fbox{\includegraphics[scale=1.4]{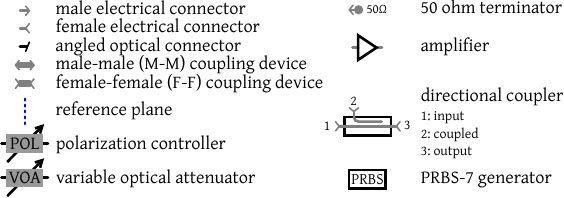}}}.

Wide band directional couplers (50 MHz to 6 GHz, type: Mini Circuits ZHDC-10-63-S+) are used to measure ``in-line'' with a fixed phase relation to the reference planes.
A directional coupler on the output is not necessarily needed, but is used in order to keep the transfer function for both reference planes to the oscilloscope as equal as possible.
Two 16 inch phase stable coaxial SMA cables (Huber+Suhner Mini141-16) are used to connect the oscilloscope.

An Ethernet transceiver evaluation board (Texas Instruments TLK1221) was used as PRBS-7 generator, together with a wide band RF amplifier (Multilink Techn. Corp. MTC5515-751) to drive the MZM (OEQuest.com LN-CHIP-12.5G).
The MZM needs a bias voltage which determines the optical phase shift for optimum  modulation.
A DFB laser at 1490 nm (Agilecom WSLS-149003C1424-20) was used for the optical delay measurements.
This laser can either be directly modulated up to 3.5 Gbit/s, or modulated by the MZM modulator\footnote{One of the measurements needs Continuous Wave (CW) light. Most SFP modules cannot be operated in CW. Without data input, the laser driver tends to oscillate and thereby modulate the laser.}. 
The polarized laser light is fed through a fibre polarization controller (Thorlabs FPC560) in order to optimize the coupling to the polarization sensitive MZM.
The delay of the optical to electrical converter can depend on the received optical power.
Therefore, the received optical power is monitored by measuring the DC current through the photo diode and held constant during the delay measurements.

Figure~\ref{fig:mzm_electrical_delay_measurement} shows the set-up to determine the electrical reference delay.
Due to the fact that SMA electrical connectors are not genderless, the EE reference delay measurement (Fig.~\ref{fig:mzm_electrical_delay_measurement}A) has two reference planes connected by a \hbox{F--F} coupling device.
An auxiliary measurement is needed to measure the delay $\Delta_{f}$ of the \hbox{F--F} coupling device (see Supplementary material) that needs to be subtracted to obtain the electrical reference delay.

The electrical delay $D_{1}$ of the MZM is determined by measurement B of Fig.~\ref{fig:mzm_electrical_delay_measurement} minus the electrical reference delay.
The optical reference delay is determined with the set-up shown in Fig.~\ref{fig:mzm_optical_delay_measurement}.
The optical delay $D_{2}$ of the MZM is determined by measurement D of Fig.~\ref{fig:mzm_optical_delay_measurement} minus the optical reference delay measurement C.
Measurement B and D are repeated in both directions of the MZM by exchanging the MZM electrical ports, respectively the optical ports.
Figure~\ref{fig:mzm_electrical_optical} shows the set-up to measure delays $D_{3}$ and $D_{4}$ (\eqref{eq:delay_eo_a} and \eqref{eq:delay_eo_b}).
$D_{4}$ is measured by changing the MZM direction (i.e. exchange the MZM ports).

The order of measurements D, E and F (Fig.~\ref{fig:mzm_optical_delay_measurement}, ~\ref{fig:mzm_electrical_optical}) is important.
Before starting measurement E, the light polarization through the MZM is optimized for maximum modulation depth with the polarization controller.
Thereafter, measurement D is performed in the same MZM direction, with the same polarization (i.e without touching the fibre circuit).
In the same manner the light polarization is optimized for measurement F and D in the other direction of the MZM.

\section{Systematic effects and uncertainties}\label{sec:systematics}
Table~\ref{tab:error_sources} is a summary of the error sources and their significance.
\begin{table}[htbp]
       \centering
       \caption{\textbf{Type B measurement uncertainties.}}
       \begin{tabular}{lrrr}
               \hline
               type B uncertainty & mean (ps) & $u_{j}$ (ps)\\
               \hline
               MZM ($\Delta EO$) & 1.1 & 1.1\\
               MZM bias voltage &  0.0 & 0.1\\
               MZM light polarization & 0.0 & 0.5\\
               RX power dependency & 0.0 & 0.1\\
               Electrical reflections & 0.0 & 0.1\\
               VOA attenuation & 0.0 & 0.3\\
               thermal laser & 0.0 & 0.4 \\
               thermal MZM & 0.0 & 0.9 \\
               thermal PIN & 0.0 & 0.1 \\
               mechanical accessory tool & 0.0 & 0.1\\
               stability & 0.0 & 0.5\\
               \hline
       \end{tabular}\label{tab:error_sources}
\end{table}

The Pockels effect in a LiNbO$_{3}$ MZM is \textit{nearly} instantaneous and the speed is governed by the photon--phonon coupling~\cite{lit:pap:PRB-schwarz-1996} through the electrically induced displacement of the atoms in the crystal.
The lowest energy resonances lie in the order of 30 cm$^{-1}$, determining the refractive index, and thus group delay for microwave signals~\cite{lit:pap:JIMTW-unferdorben-2015}.
With a separation between the electrical and optical wave guide of at most 100 $\mu$m~\cite{lit:pap:IJSTQE-wooten-2000}, and a refractive index $n=6.6$~\cite{lit:pap:JIMTW-unferdorben-2015}, the worst case EO delay is approximately $nL/c\approx2.2$ ps and we estimate this delay therefore as 1.1(1.1) ps.
It is expected that for MZMs specifically designed for this task, $\Delta EO$ can be kept small and that the device parameters needed for its determination can be accurately measured.

The MZM bias voltage is adjusted for optimal symmetric modulation.
The delay versus bias voltage around the operating point is 1.9 ps/V and contributes an uncertainty of 0.1 ps.

The maximum observed delay change due to changing light polarization over a wide range was 11.1 ps.
The polarization was optimized and controlled during the measurements with the MZM.
Under these conditions the contribution due to polarization is small (0.5 ps).

In our set-up the optical to electrical converter is a PIN diode (Terahertz Technologies Inc. TIA-1200).
The optical power versus delay dependency of the PIN diode over the range 0.6 to -17.4 dBm was determined to be 0.3 ps/dB.
The received optical power is held constant within 0.1 dB during the actual delay measurements by monitoring the DC current through the PIN diode.
Therefore, the systematic error is less than 0.1 ps.

The electrical impedance of the MZM is not perfectly matched to 50 ohms.
Electrical reflections are therefore expected to influence the delay measurements.
A maximum relative delay variation of less than 0.1 ps was measured (see Supplementary material).

Furthermore, it was checked that the delay measurements are not influenced by the Variable Optical Attenuator (VOA).
The maximum relative delay variation observed was 0.1 ps/dB. In our set-up the VOA is used over a 6 dB range (uncertainty 0.6 ps).

Thermal effects in the electronics potentially influence the delay measurements.
During multiple measurement sessions the temperature was $x\pm1$ degrees Celcius.
Delay variations due to temperature changes are evaluated to be 0.9 ps/K for the MZM modulator and 0.4 ps/K for the laser diode.
Delay variations due to temperature changes of the PIN diode was evaluated to be 0.1 ps/K.
Thermally induced delay changes of the optical fibers and microwave cables are negligible.

The length measurement of the coupling devices is described in the supplement and corresponds to a timing error of less than 0.1 ps.

It has been taken care of creating a mechanically stable set-up that is fixed onto an optical breadboard.
Multiple OE delay measurements over several weeks contribute to a spread of 1.8 ps with a $u_{i}$ of 0.5 ps.
A number of small error sources, such as electrical/optical connector reconnection and cable/fiber stability are included in table~\ref{tab:error_sources} under "stability".

\section{Experimental results}
Results for the delays A to F (Fig.~\ref{fig:mzm_electrical_delay_measurement} to ~\ref{fig:mzm_electrical_optical}) of one of the measurement series, together with $\Delta_{f}$, are shown in Table~\ref{tab:measurement_results}.

\begin{table}[htbp]
       \centering
       \caption{\textbf{Measurement A to F and $\Delta_{f}$, including type A uncertainties, of a single measurement series.}}
       \begin{tabular}{crll}
               \hline
               measurement & mean (ps) & $s_{i}$ (ps) & $v_{i}$\\
               \hline
               A &   936.5  & 0.059 &  24\\
               B &  1296.3  & 0.041 &  49\\
               C & 53086.2  & 0.228 &  24\\
               D & 71908.7  & 0.386 &  49\\
               E & 12949.8  & 0.177 &  24\\
               F & 13926.7  & 0.162 &  24\\
               $\Delta_{f}$ & 85.4 & 0.1 & 24 \\
               \hline          
       \end{tabular}\label{tab:measurement_results}
\end{table}

Delays $D_{1}$ to $D_{4}$ (see ~\eqref{eq:main_oe}) can be calculated from measurements A to F and $\Delta_{f}$ according to
\begin{subequations}
       \begin{align}
       D_{1}  &= B - (A - \Delta_{f})\label{eq:D1},\\
       D_{2}  &= D - C\label{eq:D2},\\
       D_{3}  &= E - (A - \Delta_{f})\label{eq:D3},\\
       D_{4}  &= F - (A - \Delta_{f})\label{eq:D4}.
       \end{align}
\end{subequations}
Rewriting ~\eqref{eq:main_oe} and substituting ~\eqref{eq:D1} to ~\eqref{eq:D4} yields:
\begin{equation}
    \Delta OE = \frac{-A - B + C - D + E + F + \Delta_{f}}{2} - \Delta EO \label{eq:main_oe_results}
\end{equation}
A worst case estimation of the uncertainty per measurement series A to F and $\Delta_{f}$, including systematic effects as given in Sec.~\ref{sec:systematics}, yields a standard deviation of 2 ps.
A total of 15 measurement series were performed, showing a standard deviation for $\Delta OE$ of 0.5 ps.
The optical-to-electrical delay ($\Delta OE$) is therefore determined with a conservative error as 2953(2) ps.
Knowing this delay, the EO delay of the MZM in either of it's directions is known as well.

\section{Discussion}
It is clear that the MZM used in our set-up was not optimized for the method presented. 
An optimized MZM would have physical parameters that are measurable in order to determine its $\Delta EO$ delay.
It should have a well defined, impedance matched, electrical transmission line which is accessible at both ends.

Impedance mismatch of the electrical transmission line of the MZM causes distortion of the waveforms used for cross-correlation, resulting in a timing offset.
If waveforms are recorded digitally, one could afterwards correct for impedance mismatches.
The same holds for differences in rise- and fall-times.
Future research focused on these aspects might therefore lead to smaller attainable uncertainties.

\section{Conclusion and Outlook}
We have introduced a new principle to determine the signal delay from an optical reference plane to an electrical reference plane of an opto-electrical receiver,
and thus of the electro-optical transmitter used for its determination.
We have shown a method based on this principle resulting in a $\Delta OE$ determination with 2 ps precision.
The introduced method enables cross-domain comparison of time codes and absolute calibration of optical transceivers for fibre-optic time-dissemination equipment with ps-level uncertainty.
This method can be extended to other signals, modulation types and optical line codes as long as these can be generated using MZMs.

The introduced principle enables the direct comparison of optical \emph{time} signals with their electrical counterparts.
Based on the experience gained during the presented experiment the authors think that precisions in the sub-picosecond regime are feasible.
The presented method also enables high accuracy time transfers of optical clocks and their comparison with electrical time standards.
Therefore we expect it to be an enabling technique for future generations of high accuracy time-transfer systems.


\section*{Funding Information}
Part of this work is funded by ASTERICS (https://www.asterics2020.eu/) European Commission grant no 653477.

J.C.J. Koelemeij acknowledges support from the Netherlands Organisation for Scientific Research NWO and the Technology Foundation STW.

\section*{Acknowledgments}
The authors thank Guido Visser (Nikhef) for technical support and discussions about the microwave electronics and Jan-Willem Schmelling (Nikhef) for fibre optic support. Tjeerd Pinkert lovingly thanks Esther Pinkert-Wijnbeek for their private time taken to finish this work.

\newpage
\appendix

\begin{center}
{\LARGE \tjpstitle: supplementary material}
\end{center}

\begin{abstract}%
This document provides supplementary information to ``\tjpstitle''.
In the main article a method is described to determine the absolute time delay of a waveform from an optical reference plane to an electrical reference plane and vice versa, with ps-level uncertainty.
The supplementary information provided here describes equipment, settings and boundary conditions.
Section~\ref{sec:supplement_cross-correlation} evaluates boundary conditions for interpolated cross-correlation.
In Sec.~\ref{sec:supplement_FF_measurement} and~\ref{sec:supplement_FF_measurement_helper} we provides a method for determining the delay of the female-female (\hbox{F--F}) SMA coupling device which needs to be used during the measurements described in the main article.
A method is described in Sec.~\ref{sec:electrial_reflection} to determine delay uncertainty due to electrical reflection.
In Sec.~\ref{sec:variable_optical_attenuator} we evaluate the Variable Optical Attenuator delay.
Section~\ref{sec:measurement_setup} shows the measurement setup.
Finally, in Sec.~\ref{sec:uncertainty_evaluation} the uncertainty evaluation calculation is described.
\end{abstract}

\section{Measurement of EE delays using cross \nobreak{correlation}}\label{sec:supplement_cross-correlation}
Cross correlation was done on waveforms, sampled by a 33 GHz DSAV334A Infiniium V-Series Oscilloscope.
Waveform samples were 800 kSa long and sampled at 12.5 ps intervals.
Both decreasing sample length and increasing sample interval have an effect on the correlation result.
Full and reduced sample length correlation results, as well as different sample interval correlation results were compared. 

Reduction of the waveform sample length to \textasciitilde8 kSa (a reduction factor 100) does not significantly increase the error.
Increasing the sample interval to \textasciitilde88 ps (a reduction factor 7) does not significantly increase the error.

Note that these comparisons were based on the same dataset and thus on the same oscilloscope bandwidth and time-base quality.

In general~\cite{lit:pap:ITSP-moddemeijer-1991,lit:pap:ITSP-jacovitti-1993,lit:pap:DSP-zhang-2006} show that interpolation to find the cross correlation peak is sensitive to the sampling interval to less than a few \%.
This means that for our experiment the results for measurements A to F and $\Delta e$ can be trusted up to about 0.60 ps (5\%).

\section{Measurement of the \hbox{F--F} SMA coupling \nobreak{device}}\label{sec:supplement_FF_measurement}
\begin{figure}[htbp]
       \centering
       \fbox{\includegraphics[scale=1.8]{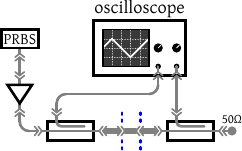}}
       \caption{electrical delay chain reference measurement (= electrical reference delay \textit{plus} the delay of an \hbox{F--F} coupling device).}\label{fig:electrical_electrical_reference}
\end{figure}

\begin{figure}[htbp]
       \centering
       \fbox{\includegraphics[scale=1.8]{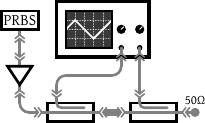}}
       \caption{auxiliary measurement to determine delay $\Delta_{e}$.}\label{fig:auxilary_measurement}
\end{figure}

The electrical delay chain measurement shown in Fig.~\ref{fig:electrical_electrical_reference}\footnote{Symbols used throughout the figures:
       
\fbox{\includegraphics[scale=1.4]{legend}}}.
includes the delay of an \hbox{F--F} coupling device. The auxiliary measurement set-up shown in Fig.~\ref{fig:auxilary_measurement} is needed to measure the delay of the \hbox{F--F} coupling device.
However, this cannot be done directly.
The difference between the delay measurements shown in Fig.~\ref{fig:electrical_electrical_reference} and ~\ref{fig:auxilary_measurement} is the delay of a \hbox{F--F} plus \hbox{M--M} (male-male) coupling device ($\Delta_{e}$).

The propagation speed of 3.5 mm coupling devices that use air as dielectric, is defined by the speed of light (c).
The \hbox{F--F} delay ($\tau_{f}$) and \hbox{M--M} delay ($\tau_{m}$) are determined by measuring their mechanical length with the aid of an accessory tool (see equation~\eqref{eq:vacuum_delay_f_f}, \eqref{eq:vacuum_delay_m_m} and Sec.~\ref{sec:supplement_FF_measurement_helper}).

The actual electrical delay $\Delta_{e}$ is \textasciitilde7 ps ($\Delta_{ex}$) longer than the sum of $\tau_{f}$ and $\tau_{m}$ (see equation ~\ref{eq:delay_excess}).
This excess delay is due to a small mechanical centre conductor support of dielectric material.
Under the assumption that both mechanical supports for \hbox{M--M} and \hbox{F--F} coupling devices are equal (i.e. use coupling devices from one manufacturer) the excess delay is equally divided over the coupling devices.
The actual electrical delay of each coupling device can thus be expressed according to equations~\eqref{eq:delay_m_m} and~\eqref{eq:delay_f_f}.
\begin{subequations}
       \begin{align}
       \tau_{f}  &= \dfrac{l_{f}}{c},\label{eq:vacuum_delay_f_f}\\
       \tau_{m}  &= \dfrac{l_{m}}{c},\label{eq:vacuum_delay_m_m}\\
       \Delta_{ex}  &= \Delta_{e} - \tau_{m} - \tau_{f},\label{eq:delay_excess}\\
       \Delta_{m} &= \tau_{m} + \dfrac{\Delta_{ex}}{2},\label{eq:delay_m_m}\\
       \Delta_{f} &= \tau_{f} + \dfrac{\Delta_{ex}}{2},\label{eq:delay_f_f}
       \end{align}
\end{subequations}

The length determination of the \hbox{M--M} and \hbox{F--F} SMA coupling devices is done using two separate mechanical length measurements.
First the length of \hbox{M--M} or \hbox{F--F} is measured using an accessory tool at each end.
Then the length of only the accessory tools clamped together is measured.
The second measurement is subtracted from the first to obtain the length of the \hbox{M--M} or \hbox{F--F} SMA coupling piece.
The precision of each of the two measurements is 0.01 mm.
Therefore the measurement of the \hbox{M--M} or \hbox{F--F} SMA coupling device has an error of 0.014 mm which corresponds to 0.045 ps.

\section{Mechanical accessory tool used to measure the mechanical length of 3.5 mm coupling devices}\label{sec:supplement_FF_measurement_helper}
\begin{figure}[htbp]
       \centering
       \fbox{\includegraphics[scale=1.8]{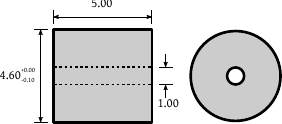}}
       \caption{mechanical accessory tool (dimensions in mm)}\label{fig:mechanical_helper_tool}
\end{figure}

The electrical reference plane of a 3.5 mm SMA connector as specified in MIL-STD-348B is mechanically difficult to access in order to enable mechanical length measurement.
The tool shown in Fig.~\ref{fig:mechanical_helper_tool} trans-locates the mechanical reference plane outside the mechanical envelope of the coupling device under test.
After inserting an accessory tool in each end of the coupling device, the total length is measured.
The actual length of the coupling device is calculated after subtraction of the length of both accessory tools.

\section{Electrical reflection}\label{sec:electrial_reflection}
\begin{figure}[htbp]
       \centering
       \fbox{\includegraphics[scale=1.8]{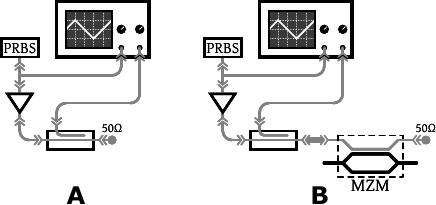}}
       \caption{set-up to measure delay influence due to electrical reflection caused by MZM impedance mismatch.}\label{fig:elec_reflection}
\end{figure}

The influence on the delay measurement of MZM electrical impedance mismatch is measured with the set-up shown in Fig.~\ref{fig:elec_reflection}.
A reference delay measurement is performed between the PRBS source and the coupled output of the 50 ohm terminated directional coupler (A).
A second measurement is performed with the MZM connected to the directional coupler while the other electrical MZM port is 50 ohm terminated (B).
The measured delay difference was less than 0.1 ps.

\section{Variable Optical Attenuator}\label{sec:variable_optical_attenuator}
The type of Variable Optical Attenuator (VOA) should be chosen with care.
Air-gap type attenuators are expected to generate variable delay as function of attenuation.
Neutral density filter type attenuators have a constant optical path length.
Therefore, in the test set-up we used the neutral density filter type attenuator to minimize delay variation.

Delay variation as function of attenuation was characterized.
Both attenuator types were places in series (Eigenlight Power Monitor 410; air-gap type and Wandel\&Goltermann OLA-25; neutral density filter type).
We measured the delay variation by changing the individual attenuators while keeping the attenuation sum constant.
Individual attenuation was changed over a range of 11 dB which resulted in a 0.9 ps delay change.
To measure the individual contribution of each attenuator was not possible.

\section{Measurement set-up}\label{sec:measurement_setup}
\begin{figure*}[htbp]
       \centering
       \fbox{\includegraphics[width=0.9\textwidth]{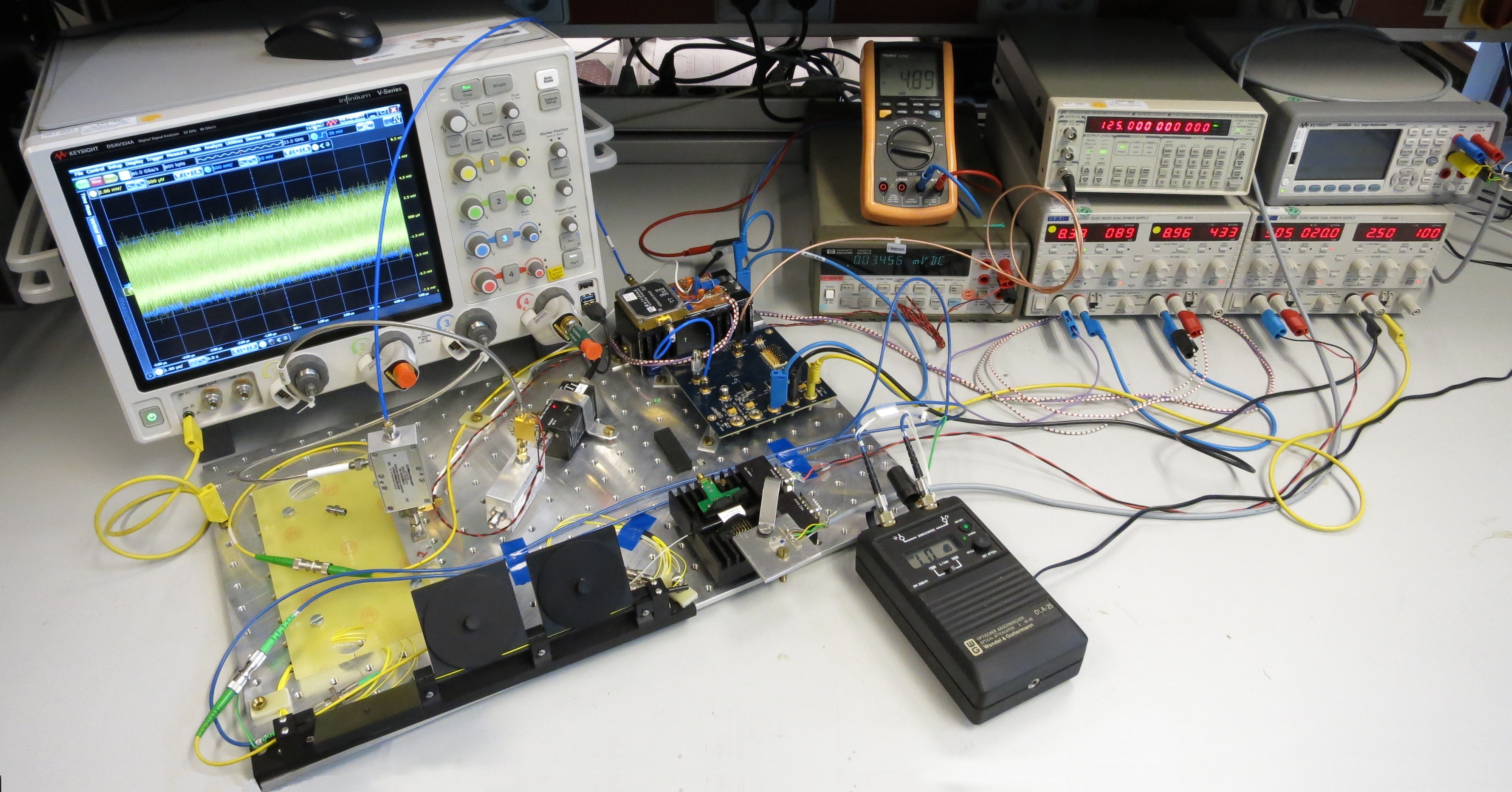}}
       \caption{measurement set-up (shown is measurement F)}\label{fig:measurement_setup_img}
\end{figure*}

\begin{figure*}[htbp]
       \centering
       \fbox{\includegraphics[width=0.9\textwidth]{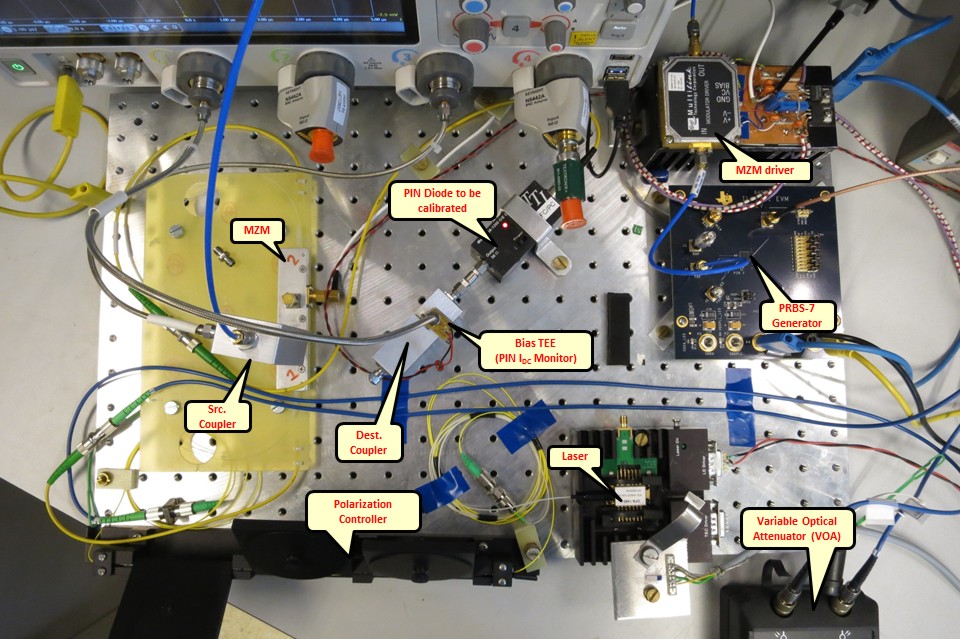}}
       \caption{detail pointing out the various components of the measurement set-up (shown is measurement F)}\label{fig:measurement_setup_detail}
\end{figure*}

Figure~\ref{fig:measurement_setup_img} shows an overview of the measurement setup.
Labels in Fig.~\ref{fig:measurement_setup_detail} show the various components in the set-up.

\section{Uncertainty evaluation}\label{sec:uncertainty_evaluation}
Equation~\ref{eq:errors_a} shows how the sum of all statistical errors is calculated.
\begin{multline}
    err_{A}=\frac{1}{2}\left\{ A_{S_{i}}^{2}+\frac{B1_{S_{i}}^{2}+B2_{S_{i}}^{2}}{2}+C_{S_{i}}^{2}\right.\\
    \left.+\frac{D1_{S_{i}}^{2}+D2_{S_{i}}^{2}}{2}+E_{S_{i}}^{2}+F_{S_{i}}^{2}+\Delta_{f_{S_{i}}}^{2} \right\}^{\frac{1}{2}} \label{eq:errors_a}
\end{multline}
Equation~\ref{eq:errors_b} shows how the sum of all errors due to systematic effects and uncertainties is calculated.
\begin{equation}
    err_{B}= \sqrt\Sigma(u_{j}^{2})\label{eq:errors_b}
\end{equation}
The total error is the combination of type A and Type B errors according to the equation
\begin{equation}
    err= (err_{A}^{2} + err_{B}^{2})^\frac{1}{2}\label{eq:errors}.
\end{equation}

%

\bibliographystyle{unsrt}
\bibliography{journalnames,literature}

\end{document}